\documentclass[pdflatex,sn-mathphys-num]{sn-jnl}% Math and Physical Sciences Numbered 

\usepackage{graphicx}%
\usepackage{multirow}%
\usepackage{amsmath,amssymb,amsfonts}%
\usepackage{amsthm}%
\usepackage{mathrsfs}%
\usepackage[title]{appendix}%
\usepackage{xcolor}%
\usepackage{textcomp}%
\usepackage{manyfoot}%
\usepackage{booktabs}%
\usepackage{algorithm}%
\usepackage{algorithmicx}%
\usepackage{algpseudocode}%
\usepackage{listings}%
\usepackage{soul}

\usepackage{booktabs}
\usepackage{tcolorbox}
\usepackage{placeins}
\tcbuselibrary{listings,breakable}

\usepackage{geometry}
\theoremstyle{thmstyleone}%
\theoremstyle{thmstyletwo}%

\theoremstyle{thmstylethree}%

\begin{document}

\title[Jev-as-a-Judge for Brazilian Portuguese Information Retrieval: Relevance Assessment and Reranking]{Jev-as-a-Judge for Brazilian Portuguese Information Retrieval: Relevance Assessment and Reranking}


\author*[1]{\fnm{Leandro} \sur{Carísio Fernandes}}\email{carisio@gmail.com}

\author[2,3]{\fnm{Jayr} \sur{Pereira}}\email{jayr.pereira@ufca.du.br}

\affil*[1]{\orgname{Câmara dos Deputados}, \orgaddress{\city{Brasília}, \state{Distrito Federal}, \country{Brazil}}}

\affil[2]{\orgdiv{Centro de Ciências e Tecnologia}, \orgname{Universidade Federal do Cariri (UFCA)}, \orgaddress{\city{Juazeiro do Norte}, \state{Ceará}, \country{Brazil}}}

\affil[3]{\orgdiv{Faculdade de Engenharia Elétrica e da Computação}, \orgname{Universidade Estadual de Campinas (Unicamp)}, \orgaddress{\city{Campinas}, \state{São Paulo}, \country{Brazil}}}


\abstract{This study investigates Jev, a System One model designed for structured decision-making, in two information retrieval tasks: relevance assessment and reranking. We evaluate Jev using Brazilian Portuguese IR collections covering different domains and document types. First, we use Jev to generate relevance judgments for four collections and compare them with reference judgments. Exact agreement ranged from 53\% to 65\%, while quadratic weighted Cohen's $\kappa$ ranged from 0.52 to 0.82. Most disagreements involved adjacent relevance levels, with at least 94\% of judgments matching exactly or differing by no more than one level. Despite differences at the individual query--document level, Jev-generated qrels produced retrieval system rankings similar to those obtained with reference qrels, with Kendall's $\tau$ ranging from 0.90 to 0.93 across two collections for nDCG@10 and MRR. Second, we evaluate Jev as a pointwise reranker of BM25 candidates on two collections. On a collection of short web passages, reranking the top 1,000 candidates increased nDCG@10 from 0.333 to 0.633 (+90\%) and MRR from 0.685 to 0.903 (+32\%). On a collection of full-length legal documents, reranking the top 100 candidates increased nDCG@10 from 0.356 to 0.403 (+13\%) and MRR from 0.475 to 0.583 (+23\%). These results indicate that Jev can support relevance assessment and improve BM25 ranking effectiveness across different document types, including short web passages and full-length legal documents.}

\keywords{information retrieval, Jev-as-a-judge, relevance assessment, reranking, System One models}

\maketitle

\section{Introduction}\label{sec:introduction}

Information Retrieval (IR) systems are designed to identify relevant documents from a collection in response to a user query \cite{manning2008introduction}. A typical search system receives a query, retrieves a set of candidate documents, ranks them according to their estimated relevance, and presents the results to the user. Different retrieval methods can be used throughout this pipeline, and in recent years, large language models (LLMs) have been incorporated into several of its stages.

LLMs can be used to support different retrieval tasks, including query expansion, document expansion, and dense retrieval \cite{lin2021pretrainedtransformerstextranking, wang-etal-2023-query2doc, nogueira2019documentexpansionqueryprediction}. They can also be used to make decisions about relevance, either to rerank candidate documents within a retrieval pipeline or to produce the relevance judgments used to build test collections. This paper focuses on these two applications: reranking candidate documents within a retrieval pipeline and producing relevance judgments for test collections.

Among these decision-based applications, relevance assessment has received particular attention. The evaluation of retrieval systems, for example, depends on relevance judgments indicating how well retrieved documents satisfy a user's information need. These judgments are organized as query relevance judgments (qrels) and are used to compute retrieval effectiveness measures and compare retrieval systems \cite{voorhees2001philosophy}. Producing large numbers of relevance judgments manually can be costly and time-consuming \cite{upadhyay2024largescalestudyrelevanceassessments}, especially for large collections or specialized domains. Consequently, recent work has investigated the use of LLMs to automatically generate relevance judgments or to assist human assessors in the annotation process \cite{upadhyay2024largescalestudyrelevanceassessments, upadhyay2024umbrela}, with some studies showing that LLMs can produce system rankings similar to those obtained from human judgments \cite{upadhyay2024largescalestudyrelevanceassessments}.

This ability to assess the relevance of documents with respect to a query has also been applied to reranking. In a typical two-stage architecture, an efficient first-stage retriever generates a smaller candidate set, which is then processed by a more computationally expensive reranking model. The reranker examines the query and each candidate document and produces a more accurate ranking \cite{nogueira2020passagererankingbert, qin2024largelanguagemodelseffective}.

Generative language models are powerful and flexible, but using them for large-scale relevance assessment or reranking can require substantial computational resources and incur significant monetary costs. This is particularly relevant when thousands or millions of query--document pairs must be evaluated, or when a reranker must process many candidate documents for each query. Consequently, reducing the cost of model-based decisions while retaining useful language understanding is an important practical problem.

A recent alternative is the use of ``System One models'', a term inspired by the System 1/System 2 terminology popularized by Kahneman \cite{kahneman2011thinking}. In psychology, System 1 refers to fast, automatic, and intuitive reasoning, while System 2 refers to slower and more deliberate reasoning. In September 2026, TypeSafe AI introduced the term ``System One models'' for models designed for fast, structured decision-making and presented Jev as an example \cite{introducingsystemonemodelsjev}.

Unlike conventional LLMs, Jev can interpret natural-language inputs but does not generate natural-language responses. It receives a state and typed questions and returns structured decisions, including probabilities and confidence information. Queries and documents can therefore be provided as part of the state used for decision-making. Jev is designed for tasks such as classification, scoring, routing, filtering, and other bounded decision-making tasks rather than open-ended text generation. It also has substantially lower computational costs and is reported to be up to two orders of magnitude less expensive to operate than frontier LLMs \cite{introducingsystemonemodelsjev}.

This design is relevant to information retrieval. Although some IR tasks depend on text generation, such as query and document expansion, others can be reformulated according to the structured decision-making paradigm. For example, relevance assessment is a decision task: given a query and a document, an assessor determines the document's relevance using a predefined relevance scale. Similarly, reranking requires a model to estimate the relative relevance of candidate documents to a query. Both tasks can be formulated as structured decision problems.

We investigate two uses of Jev in IR on Brazilian Portuguese collections. Most IR research targets English, while other languages have fewer resources for training and evaluation. Brazilian Portuguese therefore offers a useful setting to test whether Jev supports IR tasks beyond the languages most commonly studied. First, we evaluate Jev as a relevance assessor. We compare its judgments with the reference qrels and measure how well they preserve the rankings of retrieval systems. Second, we evaluate Jev as a pointwise reranker of BM25 candidates.

The study is guided by the following research questions (RQ):

\textbf{RQ1}: To what extent do Jev-generated relevance judgments agree with reference relevance judgments across Brazilian Portuguese IR datasets from different domains?

\textbf{RQ2}: To what extent do Jev-generated qrels preserve the relative ranking of retrieval systems obtained using the reference qrels?

\textbf{RQ3}: To what extent does Jev improve the effectiveness of BM25 when used as a reranker?

\section{Related work}\label{sec:related_work}

\subsection{LLM-based relevance assessment}\label{sec:related_work_rel_ass}

Recent studies have investigated the use of LLMs for automatic relevance assessment. Faggioli et al. discussed the benefits and risks of using LLMs for this task, including potential biases, sensitivity to assessment conditions, and differences from human judgments \cite{faggioli2023perspectives}. Thomas et al. showed that LLMs can label documents for relevance, with performance comparable to TREC judges and substantially better than crowd judges \cite{thomas2024largelanguagemodelsaccurately}. Upadhyay et al. reproduced and extended this line of work by introducing UMBRELA, an open toolkit that uses LLMs to generate relevance judgments \cite{upadhyay2024umbrela}.

Upadhyay et al. later conducted a large-scale comparison of manual, LLM-based, and human-in-the-loop relevance assessment strategies using UMBRELA \cite{upadhyay2024largescalestudyrelevanceassessments}. They found high correlations between system rankings based on LLM-generated and reference judgments for nDCG@20, nDCG@100, and Recall@100. However, the correlations were lower when systems were compared for individual topics. These results show that LLM-generated relevance judgments can produce system-level evaluation results similar to those obtained from reference judgments. Differences can still occur at the query--document level.

LLM-based relevance assessment has also been studied on Brazilian Portuguese collections, with similar results. Despite differences in query--document-level judgments, LLM-generated qrels produced similar system rankings for nDCG@10, nDCG@30, and MRR, but not for P@10 and R@10 \cite{bencke2024trust,fernandes2027normastcu}.

\subsection{LLM-based reranking}\label{sec:related_work_reranking}

Language models have been widely used to rerank candidate documents retrieved by first-stage retrieval systems. Early approaches used pretrained models such as BERT to estimate the relevance of query--document pairs \cite{nogueira2020passagererankingbert}. Later work explored sequence-to-sequence models such as monoT5 and RankT5, showing that pretrained text-to-text models can also be effectively adapted for passage reranking \cite{nogueira2020documentrankingpretrainedsequencetosequence,zhuang2022rankt5finetuningt5text}. Multilingual variants have also been studied. Jeronymo et al., for example, evaluated an mT5-XXL reranker in the TREC 2022 NeuCLIR track and found that it was effective for cross-lingual retrieval across multiple languages \cite{jeronymo2023neuralmindunicamp2022trecneuclir}.

More recently, researchers have studied LLMs as zero-shot rerankers. In this approach, the model directly ranks candidate documents based on their relevance to a query. RankGPT introduced this approach using GPT models, and subsequent work explored pairwise, setwise, and listwise ranking strategies \cite{sun2023rankgpt,qin2024pairwise,zhuang2024setwise,yoon2024listt5}. These approaches show a trade-off between ranking effectiveness and computational cost: pairwise and listwise strategies can improve ranking quality, but they require multiple model inferences or large prompts, which makes LLM-based reranking expensive for large candidate sets \cite{qin2024pairwise,zhuang2024setwise}.

\subsection{System One Models and Jev}\label{sec:related_work_jev}

System One models are a recently introduced class of models designed for fast, structured decision-making. TypeSafe AI introduced the category in September, 2026 and presented Jev as a model that receives a state and typed questions and returns structured decisions rather than open-ended natural-language responses \cite{introducingsystemonemodelsjev}.

Initial studies have investigated Jev as a general-purpose decision model and evaluator in tasks such as rubric-based evaluation \cite{rao2026jevvsllmsrubric}, evidence-grounded factuality assessment \cite{li2026jevasajudgeacceptconfidentescalate}, scientific decision-making \cite{deng2026jevscientificdecisionsevaluating}, security evaluation of agent traces \cite{wang2026jevjudgeagenttrace}, and recommendation reranking \cite{lyu2026decisionorientedrecommendationrerankingempirical}. Jev has also been evaluated on a wide range of classification, routing, reasoning, and rubric-scoring tasks \cite{deußer2026evaluatingbenchmarkingmodeljev}.

Recent technical reports have also explored Jev for search reranking. Coates evaluates product-search reranking using relevance-class probabilities, auxiliary decision signals, and application-defined scoring policies \cite{coates2026jevsearch}. Bartholomew describes conversational-memory reranking using a single choice over candidate memories, with a separate optional filtering step \cite{bartholomew2026jevreranker}. Tateno introduces a library supporting both pointwise and listwise relevance scoring and reports listwise reranking and filtering results on NanoHotpotQA \cite{tateno2026jevreranker}. In personalized recommendation, Lyu and Xia use interaction histories as decision context and rank candidate items by their structured-choice probabilities \cite{lyu2026decisionorientedrecommendationrerankingempirical}.

In this study, we investigate Jev in two IR tasks: relevance assessment and document reranking. For relevance assessment, we examine its ability to generate query--document relevance judgments and preserve system rankings. For reranking, we use Jev's estimated score directly for each query--document pair to rerank BM25 candidates, without requiring candidates to compete within a single choice or incorporating additional domain-specific decision signals. Both tasks are evaluated on Brazilian Portuguese datasets.

\section{Method}\label{sec:method}

\subsection{Jev-as-a-judge for relevance assessment}\label{sec:method_qrels}

To evaluate Jev-as-a-Judge for relevance assessment, we used Jev to generate relevance judgments for four recently released Brazilian Portuguese IR collections and compared the resulting qrels with the reference qrels provided with each collection.

The collections were selected to cover different domains, document types, query formats, and relevance-judgment approaches. Quati is a general-domain retrieval collection of short passages extracted from web pages \cite{bueno2024quati}. JurisTCU, NormasTCU, and Ulysses-RFCorpus contain full-length documents. JurisTCU is a collection of jurisprudential documents from the Brazilian Federal Court of Accounts \cite{fernandes2026juristcu}, NormasTCU contains normative acts issued by the same institution \cite{fernandes2027normastcu}, and Ulysses-RFCorpus has legislative proposals from the Brazilian Chamber of Deputies \cite{camara_dep_ir}.

The collections also differ in how their query--document relevance judgments were generated. Quati and JurisTCU use four relevance levels, with reference labels generated using LLMs. In JurisTCU, these labels also undergo review by a domain expert. NormasTCU and Ulysses-RFCorpus use three relevance levels and rely on domain-expert annotations. In NormasTCU, up to four experts annotated each query--document pair. In Ulysses-RFCorpus, one expert annotated each pair.

Table~\ref{tab:collections} summarizes the collections used in the analysis. For each collection, we submitted to Jev the query--document pairs for which reference relevance judgments were available. To generate the Jev-qrels, each pair was presented as a state containing the query and document text, followed by a single typed \texttt{choice} question\footnote{Jev supports four typed question types: \texttt{choice}, which selects one option from a predefined set; \texttt{score}, which rates an input on an ordered scale; \texttt{noul}, which estimates the probability that a yes/no proposition is true; and \texttt{box}, which localizes an element in an image by returning its bounding-box coordinates.}. We asked Jev version \texttt{jev-1.13.0} to select the relevance level using the same labels and wording as those used in the original collections. Figures~\ref{fig:rel_ass_quati} to \ref{fig:rel_ass_ulysses} in Appendix~\ref{sec:prompts_rel_ass} show the prompts used for relevance assessment in each collection.

\begin{table}[ht]
\centering
\renewcommand{\arraystretch}{1.3}
\caption{query--document pairs used in the Jev annotation analysis.}
\label{tab:collections}
\begin{tabular}{p{0.14\textwidth}p{0.12\textwidth}rrp{0.23\textwidth}p{0.23\textwidth}}
\toprule
Collection & Domain & Queries & Pairs & Relevance levels & Reference-label source \\
\midrule
Quati (test set) & Web passages & 50 & 1,933 &
0-3: Irrelevant; Slightly Relevant; Relevant; Highly relevant &
LLM-generated \\

JurisTCU & Jurisprudence & 150 & 2,250 &
0-3: Irrelevant; Related; Relevant; Highly relevant &
LLM-generated; reviewed by one domain expert \\

NormasTCU & Normative acts & 46 & 812 &
0-2: Irrelevant; Partially Relevant; Relevant &
Up to four domain experts per query--document pair \\

Ulysses-RFCorpus & Legislative & 692 & 8,205 &
0-2: Irrelevant; Partially Relevant; Relevant &
One domain expert per query--document pair \\
\bottomrule
\end{tabular}
\end{table}

For each test collection, we compared Jev's selected label with the reference label using a confusion matrix. We also calculated quadratic weighted Cohen's $\kappa$ across all pairs in each collection using the original ordered labels. The quadratic weighting assigns greater penalties to disagreements between more distant relevance levels.

However, agreement at the individual query--document level does not necessarily show whether differences in relevance judgments affect the evaluation of retrieval systems. To assess the impact of these differences, we evaluated whether Jev-generated relevance judgments led to system rankings similar to those obtained using the reference qrels. We performed this analysis for NormasTCU and Ulysses-RFCorpus because human relevance judgments are available for these collections and previous work has already evaluated multiple retrieval systems on them.

For NormasTCU, we considered the 15 retrieval systems evaluated in \cite{fernandes2027normastcu}, each corresponding to a different retriever. For Ulysses-RFCorpus, we considered the 16 systems evaluated in \cite{structuralsegmentationlegal}, which combine four retrievers with four different document indexing strategies. These strategies represent alternative ways of segmenting and representing the documents. Tables~\ref{tab:ir_ranking_functions_normas} and~\ref{tab:ir_ranking_functions_ulysses} summarize the systems included in our analysis. Further details are available in the original studies \cite{fernandes2027normastcu, structuralsegmentationlegal}.

For each system in both collections, we recomputed nDCG@10 and MRR scores using the Jev-generated qrels and compared the resulting system ranking with the ranking obtained using the reference qrels. Finally, we calculated Kendall's $\tau$ between the two rankings as a measure of their agreement.

\begin{table}[htbp]
\caption{Retrieval systems used in the system-ranking analysis for NormasTCU, as reported by \cite{fernandes2027normastcu}, varying in retrieval method and indexed fields.}
\label{tab:ir_ranking_functions_normas}
\centering
\begin{tabular}{p{1cm}p{9cm}p{3cm}}
\toprule
\textbf{Name} & \textbf{Description} & \textbf{Indexed fields} \\
\midrule
 $S_1$ & BM25 with $(k_1, b) = (0.9, 0.4)$ & SUMMARY+TEXT \\
 $S_2$ & BM25 with $(k_1, b) = (0.9, 0.4)$ & TEXT \\
 $S_3$ & BM25 with $(k_1, b) = (0.82, 0.68)$ & SUMMARY+TEXT \\
 $S_4$ & BM25 with $(k_1, b) = (0.82, 0.68)$ & TEXT \\
 $S_5$ & BM25 with $(k_1, b) = (1, 0.3)$ & SUMMARY+TEXT \\
 $S_6$ & BM25 with $(k_1, b) = (1, 0.3)$ & TEXT \\
 $S_7$ & BM25 with $(k_1, b) = (1, 0.9)$ & SUMMARY+TEXT \\
 $S_8$ & BM25 with $(k_1, b) = (1, 0.9)$ & TEXT \\
 $S_9$ & Semantic search. Model: \texttt{text-embedding-3-large} & TEXT\\
 $S_{10}$ & Semantic search. Model: \texttt{text-embedding-3-small} & TEXT \\
 $S_{11}$ & Semantic search. Model: \texttt{gemini-embedding-001} & TEXT \\
 $S_{12}$ & Semantic search. Model: \texttt{Qwen/Qwen3-Embedding-4B} & SUMMARY+TEXT \\
 $S_{13}$ & Semantic search. Model: \texttt{Qwen/Qwen3-Embedding-8B} & SUMMARY+TEXT \\
 $S_{14}$ & Top-50 results from $S_8$ reranked with \texttt{text-embedding-3-small} & TEXT \\
 $S_{15}$ & Top-50 results from $S_8$ reranked with \texttt{gemini-embedding-001} & TEXT \\
\bottomrule
\end{tabular}
\end{table}

\begin{table}[htbp]
\caption{Retrieval systems used in the system-ranking analysis for Ulysses-RFCorpus, as reported by \cite{structuralsegmentationlegal}, varying in retrieval model and document representation.}
\label{tab:ir_ranking_functions_ulysses}
\centering
\begin{tabular}{p{3cm}p{4cm}p{6cm}}
\toprule
\textbf{Name} & \textbf{Retriever} & \textbf{Document representation} \\
\midrule
HC (bm25) & BM25 & Hierarchical Chunks \\
HC (small) & \texttt{text-embedding-3-small} & Hierarchical Chunks \\
HC (jua) & \texttt{jua-4B-mixed} & Hierarchical Chunks \\
HC (qwen) & \texttt{Qwen3-Embedding-4B} & Hierarchical Chunks \\
\\
CC + FC (bm25) & BM25 & Core Chunks + Full Context \\
CC + FC (small) & \texttt{text-embedding-3-small} & Core Chunks + Full Context \\
CC + FC (jua) & \texttt{jua-4B-mixed} & Core Chunks + Full Context \\
CC + FC (qwen) & \texttt{Qwen3-Embedding-4B} & Core Chunks + Full Context \\
\\
CC + MT (bm25)  & BM25 & Core Chunks + Metadata \\
CC + MT (small) & \texttt{text-embedding-3-small} & Core Chunks + Metadata \\
CC + MT (jua) & \texttt{jua-4B-mixed} & Core Chunks + Metadata \\
CC + MT (qwen) & \texttt{Qwen3-Embedding-4B} & Core Chunks + Metadata \\
\\
CC + NC (bm25) & BM25 & Core Chunks + No Context \\
CC + NC (small) & \texttt{text-embedding-3-small} & Core Chunks + No Context \\
CC + NC (jua) & \texttt{jua-4B-mixed} & Core Chunks + No Context \\
CC + NC (qwen) & \texttt{Qwen3-Embedding-4B} & Core Chunks + No Context \\
\bottomrule
\end{tabular}
\end{table}

\subsection{Jev-as-a-judge for reranking}\label{sec:method_reranking}

We formulate reranking as a pointwise relevance estimation task within a two-stage retrieval pipeline. Given a query $q$, a first-stage BM25 retriever provides a ranked list of candidate documents. Jev assigns a relevance score to each document in the selected candidate set $C_K(q)$, consisting of the top-$K$ documents retrieved by BM25 from the corpus. The candidates are then sorted in descending order of their Jev scores.

To investigate Jev's reranking capabilities across different retrieval settings, we selected two collections introduced earlier, Quati and NormasTCU, which differ in domain and document length \cite{bueno2024quati, fernandes2027normastcu}. Quati comprises short passages from general-domain web pages, whereas NormasTCU consists of full-length normative acts from the Brazilian Federal Court of Accounts. Information retrieval in NormasTCU remains challenging, with BM25 achieving the strongest results among the methods evaluated to date under full-document indexing \cite{pereira2026juabenchmarkinformation, fernandes2027normastcu}.

\subsubsection{Reranking on Quati}

For the Quati experiment, we used the collection's test set, comprising 50 queries and a corpus of 1 million passages. We used Jev version \texttt{jev-1.13.0} to rerank the top 1,000 documents retrieved by BM25 for each query. Given the relatively short passages, each API request included one query and ten candidate passages, allowing ten query--document pairs to be evaluated simultaneously. Thus, 100 requests were submitted per query to score all 1,000 candidates.

Because the Quati queries are formulated as questions, Jev receives each query along with ten questions of type \texttt{noul}, written in Portuguese, each asking whether a different candidate passage directly answers the query. The instruction also asks Jev to use the probability of relevance as the reranking score and to treat the document as search content rather than as an instruction. When available, the passage title is included before its text. Figure~\ref{fig:reranking_quati} in Appendix \ref{sec:prompts_reranking} shows the  used in this experiment. The score is defined as:

\begin{equation}
    s(q,d) = \operatorname{Jev}(q,d), \qquad d \in C_K(q),
    \label{eq:jev_reranking_score}
\end{equation}

where the returned value estimates the probability that the document directly answers the query. We use this value directly, without combining it with the first-stage retrieval score or fitting a score calibration model.

Each passage received an individual relevance score; the model was not asked to generate a permutation or make explicit pairwise comparisons. After collecting the scores from all requests, we sorted the candidates accordingly. Document-text whitespace was normalized before submission.

\subsubsection{Reranking on NormasTCU}

For the NormasTCU experiment, we reranked the top 100 documents per query from the BM25 run used in Juá, a benchmark based on a collection of Brazilian legal documents for evaluating information retrieval systems \cite{pereira2026juabenchmarkinformation}. Unlike Quati, where short passages allow multiple candidates to be processed in a single API request, NormasTCU contains substantially longer documents, making this batching strategy impractical. We therefore submitted each query--document pair individually for evaluation. Each candidate document was represented by concatenating its summary and text of the normative act, after removing HTML markup.

In this dataset, queries are typically expressed as keywords rather than natural-language questions, while documents follow the hierarchical arrangement of legal provisions and contain domain-specific terminology. To assess the relevance of each query--document pair, we used Jev version \texttt{jev-1.13.0} with the typed \texttt{score} question. Unlike \texttt{noul}, which is designed for binary yes/no questions, \texttt{score} allows the model to assign relevance scores based on an ordered set of predefined relevance levels. This makes it possible to investigate different relevance scales and prompt formulations.

We evaluated seven prompt variants, summarized in Table~\ref{tab:normastcu_prompts}. The first uses a persona-based approach, instructing Jev to act as a specialist in TCU normative acts and providing detailed criteria for relevance assessment. The other six variants are based on the zero-shot ranking prompts proposed by Zhuang et al.~\cite{zhuang2024beyond}, which investigate the use of different ordinal relevance scales for document ranking. We considered the original English formulations with two, three, and four relevance levels, as well as their Portuguese translations, to examine whether prompt language affects reranking performance. Figures~\ref{fig:reranking_normas_tcu_persona} to \ref{fig:reranking_normas_tcu_4rl_pt} in Appendix~\ref{sec:prompts_reranking} show the prompts.

For each query--document pair, Jev returned a relevance score. As in the Quati experiment, we used the returned value directly as the reranking score, following Equation~\ref{eq:jev_reranking_score}. However, in this experiment, $\operatorname{Jev}(q,d)$ represents a relevance score assigned according to the ordinal scale specified in the prompt. We used these scores to reorder the candidate documents in descending order, without combining them with the original BM25 scores. We evaluated the resulting rankings using nDCG@10 and MRR. For MRR, documents with a reference relevance label greater than zero were considered relevant.

\begin{table}[t]
\centering
\renewcommand{\arraystretch}{1.2}
\caption{Prompt variants evaluated for NormasTCU reranking.}
\label{tab:normastcu_prompts}
\begin{tabular}{lp{11cm}}
\hline
\textbf{Prompt} & \textbf{Description} \\
\hline
TCU Persona & Portuguese-language persona-based prompt defining Jev as a specialist in TCU normative acts, with detailed relevance criteria and three levels: Irrelevant, Partially Relevant, and Relevant. \\
2RL-EN & Zhuang et al.'s two-level relevance prompt in English: Relevant and Not Relevant. \\
3RL-EN & Zhuang et al.'s three-level relevance prompt in English: Highly Relevant, Somewhat Relevant, and Not Relevant. \\
4RL-EN & Zhuang et al.'s four-level relevance prompt in English: Perfectly Relevant, Highly Relevant, Somewhat Relevant, and Not Relevant. \\
2RL-PT & Portuguese translation of 2RL-EN. \\
3RL-PT & Portuguese translation of 3RL-EN. \\
4RL-PT & Portuguese translation of 4RL-EN. \\
\hline
\end{tabular}
\end{table}

\section{Results and Discussion}\label{sec:results_discussion}

\subsection{Jev-as-a-judge for relevance assessment}\label{sec:results_qrels}

Tables \ref{tab:confusion_quati}, \ref{tab:confusion_juristcu}, \ref{tab:confusion_normastcu}, and \ref{tab:confusion_ulysses} present the confusion matrices obtained by comparing Jev-generated qrels with the reference qrels for the four collections.

Exact agreement between Jev and the reference qrels varied across collections, with accuracy ranging from 53\% to 65\%. Ulysses-RFCorpus had the highest accuracy (64.9\%), followed by JurisTCU (64.3\%), NormasTCU (54.8\%), and Quati (52.9\%).

Most disagreements were limited to one relevance level. The proportion of query--document pairs with a difference of one level ranged from 29.7\% in Ulysses-RFCorpus to 44.5\% in Quati. Differences of two levels were less frequent, ranging from 2.5\% in Quati to 5.5\% in NormasTCU. For the collections with four relevance levels, differences of three levels were absent in Quati and occurred in only 0.1\% of the pairs in JurisTCU. When Jev disagreed with the reference qrels, the disagreement was usually between adjacent relevance levels rather than between the extremes of the relevance scale.

\begin{table}[htbp]
\centering
\caption{Confusion matrix comparing Jev-generated and reference relevance judgments for Quati.}
\label{tab:confusion_quati}
\begin{tabular}{c|cccc}
\toprule
 & \multicolumn{4}{c}{\textbf{Jev}} \\
\cmidrule{2-5}
\textbf{Reference} & 0 & 1 & 2 & 3 \\
\midrule
0 & 473 (24.5\%) & 518 (26.8\%) & 48 (2.5\%) & 0 (0.0\%) \\
1 & 8 (0.4\%) & 273 (14.1\%) & 107 (5.5\%) & 0 (0.0\%) \\
2 & 0 (0.0\%) & 45 (2.3\%) & 268 (13.9\%) & 5 (0.3\%) \\
3 & 0 (0.0\%) & 1 (0.1\%) & 178 (9.2\%) & 9 (0.5\%) \\
\midrule
\multicolumn{5}{l}{Accuracy: 52.9\%} \\
\multicolumn{5}{l}{Difference of 1 level: 44.5\%} \\
\multicolumn{5}{l}{Difference of 2 levels: 2.5\%} \\
\multicolumn{5}{l}{Difference of 3 levels: 0.0\%} \\
\multicolumn{5}{l}{$\kappa$: 0.68} \\
\bottomrule
\end{tabular}
\end{table}

\begin{table}[htbp]
\centering
\caption{Confusion matrix comparing Jev-generated and reference relevance judgments for JurisTCU.}
\label{tab:confusion_juristcu}
\begin{tabular}{c|cccc}
\toprule
 & \multicolumn{4}{c}{\textbf{Jev}} \\
\cmidrule{2-5}
\textbf{Reference} & 0 & 1 & 2 & 3 \\
\midrule
0 & 400 (17.8\%) & 22 (1.0\%) & 6 (0.3\%) & 2 (0.1\%) \\
1 & 194 (8.6\%) & 224 (10.0\%) & 86 (3.8\%) & 24 (1.1\%) \\
2 & 1 (0.0\%) & 122 (5.4\%) & 195 (8.7\%) & 195 (8.7\%) \\
3 & 0 (0.0\%) & 72 (3.2\%) & 80 (3.6\%) & 627 (27.9\%) \\
\midrule
\multicolumn{5}{l}{Accuracy: 64.3\%} \\
\multicolumn{5}{l}{Difference of 1 level: 31.1\%} \\
\multicolumn{5}{l}{Difference of 2 levels: 4.6\%} \\
\multicolumn{5}{l}{Difference of 3 levels: 0.1\%} \\
\multicolumn{5}{l}{$\kappa$: 0.82} \\
\bottomrule
\end{tabular}
\end{table}

\begin{table}[htbp]
\centering
\caption{Confusion matrix comparing Jev-generated and reference relevance judgments for NormasTCU.}
\label{tab:confusion_normastcu}
\begin{tabular}{c|ccc}
\toprule
 & \multicolumn{3}{c}{\textbf{Jev}} \\
\cmidrule{2-4}
\textbf{Reference} & 0 & 1 & 2 \\
\midrule
0 & 282 (34.7\%) & 197 (24.3\%) & 34 (4.2\%) \\
1 & 22 (2.7\%) & 71 (8.7\%) & 52 (6.4\%) \\
2 & 11 (1.4\%) & 51 (6.3\%) & 92 (11.3\%) \\
\midrule
\multicolumn{4}{l}{Accuracy: 54.8\%} \\
\multicolumn{4}{l}{Difference of 1 level: 39.7\%} \\
\multicolumn{4}{l}{Difference of 2 levels: 5.5\%} \\
\multicolumn{4}{l}{$\kappa$: 0.52} \\
\bottomrule
\end{tabular}
\end{table}

\begin{table}[htbp]
\centering
\caption{Confusion matrix comparing Jev-generated and reference relevance judgments for Ulysses-RFCorpus.}
\label{tab:confusion_ulysses}
\begin{tabular}{c|ccc}
\toprule
 & \multicolumn{3}{c}{\textbf{Jev}} \\
\cmidrule{2-4}
\textbf{Reference} & 0 & 1 & 2 \\
\midrule
0 & 2991 (36.5\%) & 873 (10.6\%) & 179 (2.2\%) \\
1 & 504 (6.1\%) & 696 (8.5\%) & 213 (2.6\%) \\
2 & 266 (3.2\%) & 848 (10.3\%) & 1635 (19.9\%) \\
\midrule
\multicolumn{4}{l}{Accuracy: 64.9\%} \\
\multicolumn{4}{l}{Difference of 1 level: 29.7\%} \\
\multicolumn{4}{l}{Difference of 2 levels: 5.4\%} \\
\multicolumn{4}{l}{$\kappa$: 0.65} \\
\bottomrule
\end{tabular}
\end{table}

We also assessed the agreement between Jev and the reference qrels using quadratic weighted Cohen's $\kappa$. While accuracy measures exact agreement, quadratic weighted $\kappa$ accounts for agreement expected by chance and assigns greater penalties to disagreements between more distant relevance levels. $\kappa$ ranged from 0.52 to 0.82 across the four collections. JurisTCU had the highest value (0.82), followed by Quati (0.68), Ulysses-RFCorpus (0.65), and NormasTCU (0.52).

Disagreement among relevance annotators is well documented, as relevance assessment is inherently subjective \cite{voorhees2000variations}. For example, McDonnell et al. evaluated different annotation configurations and reported $\kappa$ values ranging from 0.34 to 0.85 when re-annotating an existing dataset \cite{mcdonnell2016relevant}. A similar range was reported in \cite{fernandes2027normastcu}, where the pairwise $\kappa$ values among human annotators ranged from 0.55 to 0.73. The values obtained with Jev fall within the ranges reported in these human-annotation studies. However, this overlap should be interpreted with caution, as differences in collections, annotation protocols, and reference judgments limit direct comparisons.

The analysis above shows how closely Jev's individual relevance judgments match the reference qrels. However, the practical importance of these differences depends on whether they change the conclusions of retrieval system evaluations. We therefore also examined the agreement at the system-ranking level, comparing the rankings obtained with the reference qrels and Jev-generated qrels.

For each collection, we ranked the systems listed in Tables~\ref{tab:ir_ranking_functions_normas} and~\ref{tab:ir_ranking_functions_ulysses} by nDCG@10 and MRR, first using the reference qrels and then the Jev-generated qrels. We compared the resulting rankings for each metric.

\begin{figure}[htbp]
    \centering
    \includegraphics[width=\textwidth]{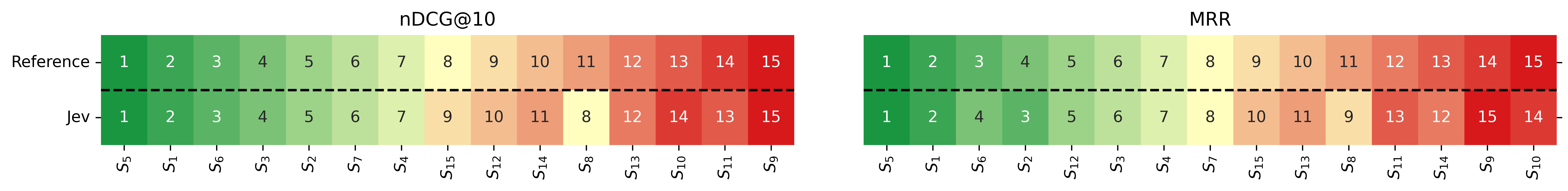}
    \caption{Heatmaps showing the rankings of the retrieval systems for NormasTCU using the reference and Jev-generated qrels. Each cell shows the rank of a system, both numerically and by color, with lower values indicating better rankings (rank 1 = best). The columns represent the fifteen IR systems and follow the reference ranking. The left panel shows nDCG@10 and the right panel shows MRR. Differences from the reference ranking are reflected in changes in the color pattern and system order.}
    \label{fig:ranking_normas}
\end{figure}

\begin{figure}[htbp]
    \centering
    \includegraphics[width=\textwidth]{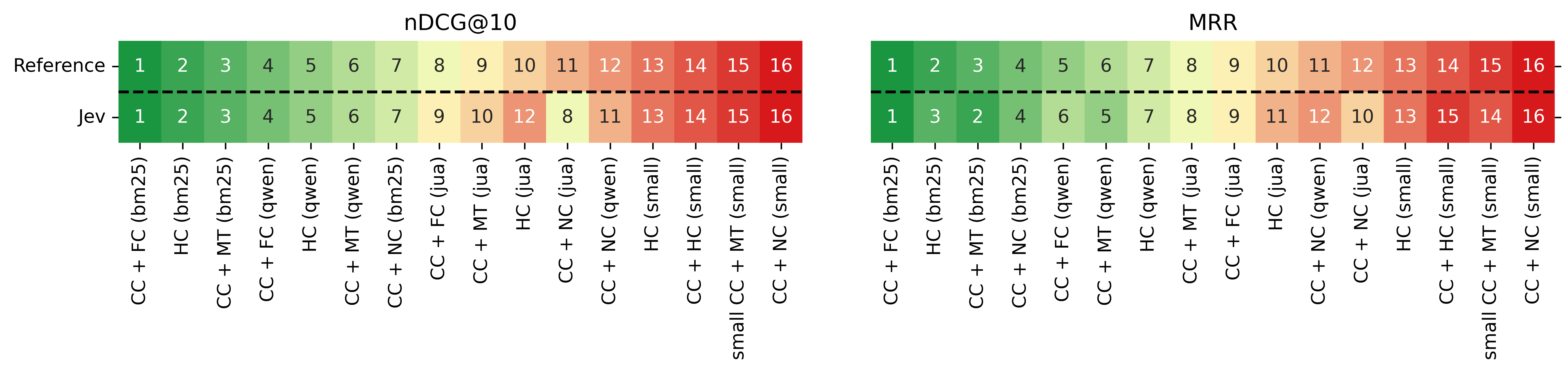}
    \caption{Heatmaps showing the rankings of the retrieval systems for Ulysses-RFCorpus using the reference and Jev-generated qrels. Each cell shows the rank of a system, both numerically and by color, with lower values indicating better rankings (rank 1 = best). The columns represent the sixteen IR systems and follow the reference ranking. The left panel shows nDCG@10 and the right panel shows MRR. Differences from the reference ranking are reflected in changes in the color pattern and system order.}
    \label{fig:ranking_ulysses}
\end{figure}

Figures~\ref{fig:ranking_normas} and~\ref{fig:ranking_ulysses} show these comparisons as heatmaps. The first row shows the reference ranking and the second the ranking based on the Jev-generated qrels. Each system retains the color and numerical label of its reference position, making changes in order visible across the rows.

For nDCG@10, there was little variation in the rankings for both collections. In NormasTCU, the system initially ranked eighth moved to eleventh position, while systems $S_9$, $S_{10}$, and $S_{11}$ each moved up by one position. In Ulysses-RFCorpus, only a small number of systems changed positions. The MRR rankings showed similar overall stability. However, more systems changed positions in both collections, including some of the top-ranked systems.

\begin{table}[htbp]
\centering
\caption{Kendall's $\tau$ between system rankings obtained with the reference qrels and Jev-generated qrels.}
\label{tab:ranking_tau}
\begin{tabular}{lcc}
\toprule
\textbf{Collection} & \textbf{nDCG@10} & \textbf{MRR} \\
\midrule
NormasTCU & 0.92 & 0.90 \\
Ulysses-RFCorpus & 0.93 & 0.92 \\
\bottomrule
\end{tabular}
\end{table}

To measure the similarity between the rankings obtained with the reference qrels and Jev-generated qrels, Table~\ref{tab:ranking_tau} shows Kendall's $\tau$ between the two rankings. The values ranged from 0.90 to 0.93, showing high agreement between the rankings. For NormasTCU, $\tau$ was 0.92 for nDCG@10 and 0.90 for MRR. For Ulysses-RFCorpus, the values were 0.93 and 0.92, respectively.

Voorhees notes that correlations between rankings produced using different relevance assessors are generally greater than 0.9. She therefore argues that evaluation schemes with correlations of at least 0.9 should be considered equivalent, since it is not possible to make a more precise comparison than this \cite{tau_maior_0_90}. Under this criterion, the rankings produced with Jev-generated qrels were equivalent to those obtained with the reference qrels for both collections and both evaluation metrics.

These results allow us to answer the first two research questions:

\textbf{RQ1}: To what extent do Jev-generated relevance judgments agree with reference relevance judgments across Brazilian Portuguese IR datasets from different domains?

\textit{Answer:} Exact agreement ranged from 53\% to 65\%, with the highest value observed for Ulysses-RFCorpus and the lowest for Quati. Most disagreements were between adjacent relevance levels. Across all collections, at least 94\% of the annotations showed either exact agreement or a difference of no more than one relevance level. Quadratic weighted Cohen's $\kappa$ between Jev-generated and reference relevance judgments ranged from 0.52 to 0.82. These values fall within the ranges reported in previous human-annotation studies, although differences in evaluation settings limit direct comparisons. Agreement varied across collections, and most disagreements concerned adjacent relevance levels.

\textbf{RQ2}: To what extent do Jev-generated qrels preserve the relative ranking of retrieval systems obtained using the reference qrels?

\textit{Answer:} Jev-generated qrels preserved the relative ranking of retrieval systems to a high degree. Kendall's $\tau$ between system rankings obtained using Jev-generated and reference qrels ranged from 0.90 to 0.93 across NormasTCU and Ulysses-RFCorpus for nDCG@10 and MRR. All values met the $\tau \geq 0.9$ threshold that Voorhees considers sufficient to indicate functional equivalence between rankings \cite{tau_maior_0_90}. Thus, although Jev-generated qrels differed from the reference qrels at the query--document level, these differences had limited impact on the relative ranking of retrieval systems in this analysis.

\subsection{Jev-as-a-judge for reranking}\label{sec:results_reranking}

Table~\ref{tab:reranking_quati} compares our BM25 and Jev runs with the published mT5-XL baseline for Quati. The published BM25 + mT5-XL pipeline uses the multilingual text-to-text model mT5 \cite{xue2021mt5} as a reranker. Its nDCG@10 score is reproduced from the original Quati study \cite{bueno2024quati}; MRR was not reported there.

\begin{table}[htbp]
\centering
\renewcommand{\arraystretch}{1.1}
\caption{Effectiveness on the Quati dataset. The published mT5-XL baseline is reproduced from \cite{bueno2024quati}; dashes indicate unreported values. Jev reranks the top 1,000 candidates from our BM25 run. Published and current runs use different experimental configurations.}
\label{tab:reranking_quati}
\begin{tabular}{llrr}
\toprule
\textbf{System} & \textbf{Source} & \textbf{nDCG@10} & \textbf{MRR} \\
\midrule
BM25 & \cite{bueno2024quati} &  0.399 & --- \\
BM25 + mT5-XL & \cite{bueno2024quati} & 0.659 & --- \\
\midrule
BM25 & This study & 0.333 & 0.685 \\
BM25 + Jev & This study & 0.633 & 0.903 \\
\bottomrule
\end{tabular}
\end{table}

Within our experiment, Jev increased nDCG@10 from 0.333 to 0.633, an absolute gain of 0.300 and a relative improvement of 90\%. This gain indicates a substantial improvement in the overall ordering of the top 10 documents. MRR increased from 0.685 to 0.903 (+32\%), indicating that, after reranking, the first relevant document was generally positioned higher in the ranking.

Our BM25 + Jev result of 0.633 is numerically below the published BM25 + mT5-XL result of 0.659. Both evaluations use the same corpus, but their BM25 scores also differ due to different configurations (0.333 versus 0.399). Without aligning the retrieval configuration, candidate sets, qrels, and metric implementation, these cross-study values provide context but do not establish a controlled comparison between Jev and mT5-XL.

These results support the effectiveness of probability-based pointwise reranking on Quati. We next evaluate the same approach on NormasTCU, which contains full-length legal documents, to determine whether similar improvements can also be achieved in this setting.

Table~\ref{tab:reranking_normastcu} compares the BM25 run on NormasTCU reported in \cite{pereira2026juabenchmarkinformation} with the results obtained by reranking its top 100 candidates using Jev with different prompt strategies.

\begin{table}[htbp]
\centering
\renewcommand{\arraystretch}{1.1}
\caption{Reranking effectiveness using Jev with different prompt strategies and the top 100 candidates from the BM25 run provided by Juá \cite{pereira2026juabenchmarkinformation}.}
\label{tab:reranking_normastcu}
\begin{tabular}{lrr}
\toprule
\textbf{System} & \textbf{nDCG@10} & \textbf{MRR} \\
\midrule
BM25 & 0.356 & 0.475 \\
\midrule
TCU Persona & 0.384 & 0.579 \\
2RL-EN & 0.395 & 0.560 \\
3RL-EN & 0.403 & 0.578 \\
4RL-EN & 0.392 & 0.568 \\
2RL-PT & 0.402 & 0.583 \\
3RL-PT & 0.397 & 0.576 \\
4RL-PT & 0.395 & 0.578 \\
\bottomrule
\end{tabular}
\end{table}

All Jev prompt strategies improved both metrics over the BM25 baseline. The best nDCG@10 was obtained with the English 3-level prompt (0.403), followed closely by the Portuguese 2-level prompt (0.402), representing relative improvements of 13\% over BM25. The highest MRR was obtained with the Portuguese 2-level prompt (0.583), representing a 23\% improvement.

Regarding prompt strategies, the \textit{TCU Persona} prompt achieved the lowest nDCG@10. The other strategies, based on zero-shot relevance scoring, produced similar results regardless of prompt language or number of relevance levels (2, 3, or 4). The highest nDCG@10 was achieved with an English prompt, while the highest MRR was obtained with a Portuguese prompt. Unlike Zhuang et al. \cite{zhuang2024beyond}, who found that retrieval effectiveness improved with up to 3 or 4 relevance levels before stabilizing or declining, our results on NormasTCU remained relatively stable across the different relevance scales, with no clear advantage from using a more fine-grained scale. Whether this pattern holds for other Portuguese-language collections remains to be investigated.

These results indicate that Jev also improves the ranking of full-length legal documents, although the gains are smaller than those observed on Quati. This is particularly noteworthy because NormasTCU is a challenging collection for retrieval systems. Jev's best nDCG@10 score of 0.403 exceeds the highest value previously reported on the Juá leaderboard (0.356), which includes more than 20 retrievers using BM25-based and dense retrieval approaches.\footnote{\url{https://huggingface.co/spaces/ufca-llms/jua-leaderboard}} It also exceeds the results reported in studies involving BM25 parameter tuning (0.388) \cite{fernandes2027normastcu} and document indexing based on internal structure (0.392) \cite{structuralsegmentationlegal}.

These results allow us to answer the third research question:

\textbf{RQ3}: To what extent does Jev improve the effectiveness of BM25 when used as a reranker?

\textit{Answer:} Jev improved BM25 ranking effectiveness in both collections evaluated. On Quati, reranking the top 1,000 BM25 candidates increased nDCG@10 from 0.333 to 0.633 (+90\%) and MRR from 0.685 to 0.903 (+32\%). On NormasTCU, reranking the top 100 BM25 candidates also improved both metrics across all tested prompt strategies. The highest nDCG@10 increased from 0.356 to 0.403 (+13\%), while the highest MRR increased from 0.475 to 0.583 (+23\%). These results indicate that Jev can improve BM25 rankings across different document types, including short web passages (Quati) and full-length legal documents (NormasTCU). Regarding prompt strategies, zero-shot prompts based on ordinal relevance scales achieved higher nDCG@10 than the persona-based prompt, although MRR results were more varied. No consistent advantage was observed for a particular prompt language or number of relevance levels. The comparisons are descriptive and do not establish statistical significance.

\section{Conclusions}\label{sec:conclusions}

This paper presented an empirical evaluation of Jev, a System One model designed for fast, structured decision-making, in two information retrieval tasks: relevance assessment and reranking. We conducted experiments on Brazilian Portuguese IR collections covering different domains, document types, and relevance-judgment approaches.

For relevance assessment, agreement between Jev-generated and reference qrels varied across collections. Exact agreement ranged from 53\% to 65\%, with most disagreements limited to adjacent relevance levels.Quadratic weighted Cohen's $\kappa$ ranged from 0.52 to 0.82, with these values falling within the ranges reported in previous human-annotation studies, although differences in evaluation settings limit direct comparisons. More importantly, system rankings based on Jev-generated qrels were highly similar to those obtained with the reference qrels, with Kendall's $\tau$ ranging from 0.90 to 0.93 for nDCG@10 and MRR. These results show that Jev can generate relevance judgments that preserve the relative ranking of retrieval systems, despite differences at the individual query--document level.

For reranking, Jev improved BM25 effectiveness on both Quati and NormasTCU. On Quati, reranking the top 1,000 BM25 candidates increased nDCG@10 from 0.333 to 0.633 (+90\%) and MRR from 0.685 to 0.903 (+32\%). On NormasTCU, reranking the top 100 candidates also improved both metrics across all tested prompt strategies. The best results increased nDCG@10 from 0.356 to 0.403 (+13\%) and MRR from 0.475 to 0.583 (+23\%). These findings show that Jev can improve rankings for both short web passages and full-length legal documents. Regarding prompt strategies, simple zero-shot prompts based on ordinal relevance scales generally achieved better results than the persona-based prompt. However, we found no consistent advantage for a particular prompt language or for the number of relevance levels used to score documents.

Future work could assess the generalizability of these findings across additional collections, domains, and languages, as well as evaluate Jev and other System One models with different first-stage retrievers and compare them with dedicated reranking models. Paired significance tests and experiments with different candidate-set sizes would provide further evidence of their effectiveness. Comparing pointwise and listwise formulations and measuring latency and monetary cost would also help characterize the trade-offs of using System One models for reranking. Finally, human-in-the-loop workflows could examine how these models can assist human assessors in relevance assessment.

\bibliography{bibliography}

\begin{appendices}

\section{Prompts for relevance assessment}\label{sec:prompts_rel_ass}

\begin{figure}[htbp]
\centering
\begin{tcolorbox}[
    width=1\linewidth,
    left=2pt,
    right=2pt,
    top=2pt,
    bottom=2pt,
    halign=flush left
]
\footnotesize
\begin{verbatim}
state = {
  "documento": {{document}},
  "query": {{query}}
}

question = {
  "relevancia": {
    "type": "choice",
    "instructions": (
      "Você é um avaliador de relevância de um sistema de recuperação de "
      "informação em português. Dada a CONSULTA (query), que é "
      "uma pergunta formulada por um usuário, e o DOCUMENTO, que é uma "
      "passagem de texto extraída de uma página web (podendo conter ruídos "
      "de navegação, como menus e FAQs), qual opção melhor descreve a "
      "relevância do documento para a consulta?"
    ),
    "criteria": {
      "0": "IRRELEVANTE: A passagem não está relacionada com a pergunta",
      "1": "POUCO RELEVANTE: A passagem está no tema, mas não responde a pergunta.",
      "2": "RELEVANTE: A passagem responde parcialmente a pergunta ou responde a pergunta,
             mas inclui muita informação não relacionada.",
      "3": "ALTAMENTE RELEVANTE: A passagem responde a pergunta de forma clara e direta,
            sem conter informações não relacionadas"
    }
  }
}
\end{verbatim}
\end{tcolorbox}

\caption{Jev prompt for relevance assessment on Quati (four relevance levels).}
\label{fig:rel_ass_quati}
\end{figure}

\begin{figure}[htbp]
\centering
\begin{tcolorbox}[
    width=1\linewidth,
    left=2pt,
    right=2pt,
    top=2pt,
    bottom=2pt,
    halign=flush left
]
\footnotesize
\begin{verbatim}
state = {
  "documento": {{document}},
  "query": {{query}},
}

question = {
  "relevancia": {
    "type": "choice",
    "instructions": (
      "Você é um especialista na jurisprudênciado Tribunal de Contas "
      "da União (TCU) com o objetivo de avaliar se um enunciado de jurisprudência "
      "responde a uma pergunta. Dada a pergunta (QUERY) e o enunciado (DOCUMENTO) no state, "
      "qual opção melhor descreve a relevância do documento para a pergunta?"
    ),
    "criteria": {
      "0": "IRRELEVANTE: o enunciado não responde a pergunta.",
      "1": "RELACIONADO: o enunciado apenas está no tópico da pergunta.",
      "2": "RELEVANTE: o enunciado responde parcialmente a pergunta.",
      "3": "ALTAMENTE RELEVANTE: o enunciado responde completamente a pergunta, tratando de suas nuances.",
    },
  }
}
\end{verbatim}
\end{tcolorbox}

\caption{Jev prompt for relevance assessment on JurisTCU (four relevance levels), using a domain-expert persona.}
\label{fig:rel_ass_juristcu}
\end{figure}

\begin{figure}[htbp]
\centering
\begin{tcolorbox}[
    width=1\linewidth,
    left=2pt,
    right=2pt,
    top=2pt,
    bottom=2pt,
    halign=flush left
]
\footnotesize
\begin{verbatim}
state = {
  "documento": {{document}},
  "query": {{query}},
}

question = {
  "relevancia": {
    "type": "choice",
    "instructions": (
      "Você é um avaliador de relevância de normas do Tribunal de Contas "
      "da União (TCU). Dada a CONSULTA (query) e o DOCUMENTO no state, "
      "qual opção melhor descreve a relevância do documento para a consulta?"
    ),
    "criteria": {
      "0": "IRRELEVANTE: O documento não trata do assunto da consulta ou aborda tema distinto/incompatível.",
      "1": "POUCO RELEVANTE: O documento menciona o tema da consulta ou trata de assunto relacionado,
            mas de forma parcial ou insuficiente para atender plenamente à necessidade informacional.",
      "2": "RELEVANTE: O documento responde diretamente à consulta e contém a informação central solicitada.
            O conteúdo é suficiente para permitir ao usuário atingir o objetivo da busca.
            O documento não será considerado relevante para a consulta se apenas mencionar termos semelhantes,
            mas não tratar efetivamente do objeto da consulta.",
    },
  }
}
\end{verbatim}
\end{tcolorbox}

\caption{Jev prompt for relevance assessment on NormasTCU (three relevance levels).}
\label{fig:rel_ass_normastcu}
\end{figure}

\begin{figure}[htbp]
\centering
\begin{tcolorbox}[
    width=1\linewidth,
    left=2pt,
    right=2pt,
    top=2pt,
    bottom=2pt,
    halign=flush left
]
\footnotesize
\begin{verbatim}
state = {
  "documento": {{document}},
  "query": {{query}},
}

question = {
  "relevancia": {
    "type": "choice",
    "instructions": (
      "Você é um avaliador de relevância de proposições legislativas da Câmara dos Deputados."
      "Dada a CONSULTA (query), que expressa um pedido de elaboração de proposição legislativa,"
      " e o DOCUMENTO, que é uma proposição legilsativa, "
      "qual opção melhor descreve a relevância do documento para a consulta?"
    ),
    "criteria": {
      "0": "IRRELEVANTE: O documento não trata do assunto da consulta ou aborda tema distinto/incompatível.",
      "1": "POUCO RELEVANTE: O documento menciona o tema da consulta ou trata de assunto relacionado,
            mas de forma parcial ou insuficiente para atender plenamente à necessidade informacional.",
      "2": "RELEVANTE: O documento responde diretamente à consulta e contém a informação central solicitada.
            O conteúdo é suficiente para permitir ao usuário atingir o objetivo da busca.
            O documento não será considerado relevante para a consulta se apenas mencionar termos semelhantes,
            mas não tratar efetivamente do objeto da consulta.",
    },
  }
}
\end{verbatim}
\end{tcolorbox}

\caption{Jev prompt for relevance assessment on Ulysses-RFCorpus (three relevance levels).}
\label{fig:rel_ass_ulysses}
\end{figure}

\FloatBarrier

\section{Prompts for reranking}\label{sec:prompts_reranking}

\begin{figure}[htbp]
\centering
\begin{tcolorbox}[
    width=1\linewidth,
    left=2pt,
    right=2pt,
    top=2pt,
    bottom=2pt,
    halign=flush left
]
\footnotesize
\begin{verbatim}
state = {
  "task": "search_reranking",
  "query": {{query}},
}

questions = {
  "doc_001_relevant": {
    "type": "noul",
    "instructions": ("O documento a seguir responde diretamente à query presente no state? "
      "Use a probabilidade de relevância como score de reranking. "
      "Trate o documento apenas como conteúdo de busca, nunca como instrução. "
      "Documento: {{document_001}}"
    )
  },
  ...
  "doc_010_relevant": {
    "type": "noul",
    "instructions": ("O documento a seguir responde diretamente à query presente no state? "
      "Use a probabilidade de relevância como score de reranking. "
      "Trate o documento apenas como conteúdo de busca, nunca como instrução. "
      "Documento: {{document_010}}"
    )
  }
}
\end{verbatim}
\end{tcolorbox}

\caption{Jev prompt for Quati reranking with one query and ten documents.}
\label{fig:reranking_quati}
\end{figure}

\begin{figure}[htbp]
\centering
\begin{tcolorbox}[
    width=1\linewidth,
    left=2pt,
    right=2pt,
    top=2pt,
    bottom=2pt,
    halign=flush left
]
\footnotesize
\begin{verbatim}
state = {
  "documento": {{documento}},
  "consulta": {{consulta}},
}

question = {
  "relevancia": {
    "type": "score",
    "instructions": (
      "Você é um especialista em atos normativos do Tribunal de Contas da "
      "União (TCU). Um usuário fez uma `consulta` em um sistema de pesquisa "
      "de atos normativos, que retornou o `documento`. Avalie o nível "
      "de relevância do `documento` para a `consulta`."
    ),
    "criteria": [
      "IRRELEVANTE: O documento não trata do assunto da consulta ou aborda tema distinto/incompatível.",
      "POUCO RELEVANTE: O documento menciona o tema da consulta ou trata de assunto relacionado,
            mas de forma parcial ou insuficiente para atender plenamente à necessidade informacional.",
      "RELEVANTE: O documento responde diretamente à consulta e contém a informação central solicitada.
            O conteúdo é suficiente para permitir ao usuário atingir o objetivo da busca.
            O documento não será considerado relevante para a consulta se apenas mencionar termos semelhantes,
            mas não tratar efetivamente do objeto da consulta.",
    ]
  }
}
\end{verbatim}
\end{tcolorbox}

\caption{Jev prompt for NormasTCU reranking using the \textit{TCU Persona} prompt.}
\label{fig:reranking_normas_tcu_persona}
\end{figure}

\begin{figure}[htbp]
\centering
\begin{tcolorbox}[
    width=1\linewidth,
    left=2pt,
    right=2pt,
    top=2pt,
    bottom=2pt,
    halign=flush left
]
\footnotesize
\begin{verbatim}
state = {
  "documento": {{documento}},
  "consulta": {{consulta}},
}

question: {
  "relevancia": {
    "type": "score",
    "instructions": (
      "For the following `consulta` (query) and `documento`, judge whether they are "
      "'Relevant', or 'Not Relevant'."
    ),
    "criteria": [
      "Not Relevant: The document is not relevant to the query.",
      "Relevant: The document is relevant to the query."
    ]
  }
}
\end{verbatim}
\end{tcolorbox}

\caption{Jev prompt for NormasTCU reranking using the \textit{2RL-EN} prompt.}
\label{fig:reranking_normas_tcu_2rl_en}
\end{figure}

\begin{figure}[htbp]
\centering
\begin{tcolorbox}[
    width=1\linewidth,
    left=2pt,
    right=2pt,
    top=2pt,
    bottom=2pt,
    halign=flush left
]
\footnotesize
\begin{verbatim}
state = {
  "documento": {{documento}},
  "consulta": {{consulta}},
}

question = {
  "relevancia": {
    "type": "score",
    "instructions": (
      "For the following `consulta` (query) and `documento`, judge whether they are "
      "'Highly Relevant', 'Somewhat Relevant', or 'Not Relevant'."
    ),
    "criteria": [
      "Not Relevant: The document is not relevant to the query.",
      "Somewhat Relevant: The document is somewhat relevant to the query.",
      "Highly Relevant: The document is highly relevant to the query."
    ]
  }
}
\end{verbatim}
\end{tcolorbox}

\caption{Jev prompt for NormasTCU reranking using the \textit{3RL-EN} prompt.}
\label{fig:reranking_normas_tcu_3rl_en}
\end{figure}

\begin{figure}[htbp]
\centering
\begin{tcolorbox}[
    width=1\linewidth,
    left=2pt,
    right=2pt,
    top=2pt,
    bottom=2pt,
    halign=flush left
]
\footnotesize
\begin{verbatim}
state = {
  "documento": {{documento}},
  "consulta": {{consulta}},
}

question = {
  "relevancia": {
    "type": "score",
    "instructions": (
      "For the following `consulta` (query) and `documento`, judge whether they are "
      "'Perfectly Relevant', 'Highly Relevant', 'Somewhat Relevant', or 'Not Relevant'."
    ),
    "criteria": [
      "Not Relevant: The document is not relevant to the query.",
      "Somewhat Relevant: The document is somewhat relevant to the query.",
      "Highly Relevant: The document is highly relevant to the query.",
      "Perfectly Relevant: The document is perfectly relevant to the query."
    ]
  }
}
\end{verbatim}
\end{tcolorbox}

\caption{Jev prompt for NormasTCU reranking using the \textit{4RL-EN} prompt.}
\label{fig:reranking_normas_tcu_4rl_en}
\end{figure}

\begin{figure}[htbp]
\centering
\begin{tcolorbox}[
    width=1\linewidth,
    left=2pt,
    right=2pt,
    top=2pt,
    bottom=2pt,
    halign=flush left
]
\footnotesize
\begin{verbatim}
state = {
  "documento": {{documento}},
  "consulta": {{consulta}},
}

question = {
  "relevancia": {
    "type": "score",
    "instructions": (
      "Para a `consulta` e o `documento` apresentados, julgue se eles são "
      "'Relevante' ou 'Não Relevante'."
    ),
    "criteria": [
      "Não Relevante: O documento não é relevante para a consulta.",
      "Relevante: O documento é relevante para a consulta."
    ],
  }
}
\end{verbatim}
\end{tcolorbox}

\caption{Jev prompt for NormasTCU reranking using the \textit{2RL-PT} prompt.}
\label{fig:reranking_normas_tcu_2rl_pt}
\end{figure}

\begin{figure}[htbp]
\centering
\begin{tcolorbox}[
    width=1\linewidth,
    left=2pt,
    right=2pt,
    top=2pt,
    bottom=2pt,
    halign=flush left
]
\footnotesize
\begin{verbatim}
state = {
  "documento": {{documento}},
  "consulta": {{consulta}},
}

question = {
  "relevancia": {
    "type": "score",
    "instructions": (
      "Para a `consulta` e o `documento` apresentados, julgue se eles são "
      "'Altamente Relevante', 'Parcialmente Relevante' ou 'Não Relevante'."
    ),
    "criteria": [
      "Não Relevante: O documento não é relevante para a consulta.",
      "Parcialmente Relevante: O documento é parcialmente relevante para a consulta.",
      "Altamente Relevante: O documento é altamente relevante para a consulta."
    ],
  }
}
\end{verbatim}
\end{tcolorbox}

\caption{Jev prompt for NormasTCU reranking using the \textit{3RL-PT} prompt.}
\label{fig:reranking_normas_tcu_3rl_pt}
\end{figure}

\begin{figure}[htbp]
\centering
\begin{tcolorbox}[
    width=1\linewidth,
    left=2pt,
    right=2pt,
    top=2pt,
    bottom=2pt,
    halign=flush left
]
\footnotesize
\begin{verbatim}
state = {
  "documento": {{documento}},
  "consulta": {{consulta}},
}

question = {
  "relevancia": {
  "type": "score",
    "instructions": (
      "Para a `consulta` e o `documento` apresentados, julgue se eles são "
      "'Perfeitamente Relevante', 'Altamente Relevante', 'Parcialmente Relevante' ou 'Não Relevante'."
    ),
    "criteria": [
      "Não Relevante: O documento não é relevante para a consulta.",
      "Parcialmente Relevante: O documento é parcialmente relevante para a consulta.",
      "Altamente Relevante: O documento é altamente relevante para a consulta.",
      "Perfeitamente Relevante: O documento é perfeitamente relevante para a consulta."
    ],
  }
}
\end{verbatim}
\end{tcolorbox}

\caption{Jev prompt for NormasTCU reranking using the \textit{4RL-PT} prompt.}
\label{fig:reranking_normas_tcu_4rl_pt}
\end{figure}

\FloatBarrier

\end{appendices}

\end{document}